\documentstyle[psfig]{mn}

\newcommand{\nc}{\newcommand}
\nc{\be}{\begin{eqnarray}}
\nc{\ee}{\end{eqnarray}}
\nc{\ng}{non-Gaussianity }
\nc{\ngn}{non-Gaussian }
\nc{\prd}{Phys. Rev. D}
\nc{\prl}{Phys. Rev. Lett.}
\nc{\apj}{ApJ}
\nc{\mnras}{MNRAS}
\nc{\physrep}{Phys. Rep.}
\nc{\nat}{Nature}
\nc{\araa}{ARA\&A}

\title{A new statistic for picking out Non-Gaussianity in the CMB}
\author[A. M. Lewin et al.]{Alex Lewin\thanks{a.m.lewin@ic.ac.uk},
Andreas Albrecht\thanks{a.albrecht@ic.ac.uk} 
and Jo\~{a}o Magueijo\thanks{j.magueijo@ic.ac.uk} \\ 
Blackett Laboratory, Imperial College,
Prince Consort Road, London SW7 2BZ, UK}

\begin{document}
\maketitle
\begin{abstract}
In this paper we propose a new statistic capable of detecting
non-Gaussianity in the CMB. The statistic is defined in Fourier space,
and therefore naturally separates angular scales. It consists of taking
another Fourier transform, in angle, over the Fourier modes within a 
given ring of scales. Like other Fourier space statistics, our statistic
outdoes more conventional methods when faced with combinations of 
Gaussian processes (be they noise or signal) and a non-Gaussian signal 
which dominates only on some scales. 
However, unlike previous efforts along these lines, our statistic is 
successful in recognizing multiple non-Gaussian patterns in a single field.
We discuss various applications, in which the Gaussian component may be
noise or primordial signal, and the non-Gaussian component may be a cosmic
string map, or some geometrical construction mimicking, say, small scale
dust maps.
\end{abstract}
\begin{keywords}
cosmic microwave background -- methods: statistical
\end{keywords}

\section{Introduction}
Current theories of structure formation may be roughly divided
into two classes: active and passive perturbations. 
According to the inflationary paradigm, quantum
fluctuations in the very early universe are produced during a period of
inflation \cite{cosmcross} and grow to 
become classical density perturbations \cite{lidlyth,bardeen}. These
perturbations evolve linearly until late times when the overdensities
become galaxies. Perturbations due to inflation are called passive
because they are seeded at some initial time near the Planck time and
then evolve `deterministically', or linearly. They leave their 
imprint on the CMB at last scattering 
\cite{pyu,nature,HS1,HS2,richard,1.3}. 
Although this is not strictly necessary, in most cases  the
fluctuations in the CMB temperature due to inflationary perturbations
form a Gaussian random field \cite{bondefst,bardstat}.

There is another class of theories of structure formation, topological
defects caused by phase transitions in the early universe \cite{Kib,vv}.
Perturbations caused by defects are known as active perturbations 
\cite{mafc,coher}
since they are continually being seeded by an evolving network of defects
through the history of the universe. The fluctuations in the CMB in
defect models have been found to be non-Gaussian \cite{phases},
even though the extent and strength of this non-Gaussianity is still far
from clear.  Recently a wide class of defect models have been shown to
be in conflict with current data \cite{neil,abr,abr2} but some viable
active models still remain \cite{mimic,durrer,abrw}. In fact, some of the
most interesting \cite{abr3,shellard} of these (based on cosmic
strings) require a non-zero cosmological constant of the sort that
currently favoured by supernova experiments \cite{sn1}.

Thus it is important to be able to distinguish Gaussian from \ngn
fluctuations. Many different tests are being tried 
\cite{Kogut,XLuo,fermag,cumul,gorski,sergei}, adapted to different
experimental settings, and types of signal. 
Our statistic is designed to be used for small fields where one or more
distinctive shapes are obscured by an extra 
Gaussian component, and so are not
visible in real space. It is our experience in previous work
\cite{fermag} that in such situations standard statistics, based in
real space, fail to recognize the non-Gaussianity of the signal.
The idea is to study the statistical properties
of map derivatives which are only sensitive to a given scale. 
In this way we may separate out different scales,
some of which may have Gaussian or nearly Gaussian fluctuations, some
very non-Gaussian. Although this scale filtering may be achieved
using the wavelet transform \cite{cumul,pando}, 
in this paper we choose to use the Fourier transform as our scale filter. 

One problem with this approach is that the Fourier transform is 
a global transformation, and therefore can only recognize 
non-Gaussian structures globally. 
It may therefore offer a rather contorted description
for complicated networks made up of essentially 
simple objects. We will however perform
a transformation over the Fourier modes: a second Fourier transform,
in angle, for modes in each ring in Fourier space. 
We will argue that by doing so we are able to recognize mostly
features of individual objects, and so bypass this problem.
This operation returns a set of quantities which are blind to
the random orientations of individual objects. Unfortunately
their random positions still affect our statistic, so there will
be a limit (albeit less stringent) 
on the number of objects in the field before our
statistic becomes confused.

The plan of this paper is as follows. 
In Section II we look at some of the ideas involved in thinking
about non-Gaussianity. In Section III we define our statistic 
and look at some of its properties and motivations.
In Section IV we apply our statistic to practical situations
in which subtle non-Gaussian signals are present. We consider 
non-Gaussian signals corrupted by the presence of  
a Gaussian process, which can be noise or primordial
signal, and which dominates on all but a narrow band of scales.
We consider two types of non-Gaussian signal. We consider
CMB maps obtained from string simulations, implementing the
algorithms in \cite{av}. We also consider geometrical constructions
 mimicking, say, small scale dust maps. 
We show how such maps look very Gaussian, 
but fail to confuse our statistic.

\section{Non-Gaussian Statistics}\label{ng}

The fluctuations in the temperature of the radiation are described by 
 $\frac{\Delta T}{T}({\bmath n})$ on a two-sphere.  Generally this is 
expanded in
spherical harmonics, but for small fields it can be expanded in
Fourier modes:
\begin{equation}\label{fourier}
  \frac{\Delta T}{T}({\bmath x})
={\int {{\rm d}^2{\bmath k}\over 2\pi}a({\bmath k}){\rm e}^{{\rm i}\bmath {k\cdot x}}}
\end{equation} 

The assumption of homogeneity and isotropy leads to the relation
between Fourier modes
\begin{equation}\label{homiso}
\langle a^*({\bmath k_1})a({\bmath k_2})\rangle=\delta({\bmath k_1}-
{\bmath k_2})P(k_2)
\end{equation} 
where $P(k)={\langle |a({\bmath k})|^2\rangle}_{|{\bmath k}|=k}$ is the
power spectrum and the averages are ensemble averages.

Homogeneity means that $\langle |\frac{\Delta T}{T}({\bmath x})|^2 \rangle = 
\langle |\frac{\Delta T}{T}({\bmath {x+b}})|^2 \rangle$ for all vectors
${\bmath b}$ which means 
$\langle a^*({\bmath k}_1)a({\bmath k}_2)\rangle = 0$ for ${\bmath k}_1 
\neq {\bmath k}_2$.  
Isotropy in real space corresponds to
isotropy in Fourier space so the power spectrum can only depend on the
modulus of ${\bmath k}$.

It is often assumed that the Fourier modes form a Gaussian random field.
The real and imaginary parts of the Fourier modes of a Gaussian random field
are drawn
from a Normal distribution with mean zero and variance $P(k)/2$ which
means that the phases are random and the moduli have a $\chi ^2$
distribution.
From our experience, we know that in real space a Gaussian random
field has little structure. We may of course have a field that obeys equation
(\ref{homiso}) but which clearly has structure in real space, for
example a field containing squares at random positions and
orientations. This is clearly non-Gaussian. 
The important point is that the averages in equation (\ref{homiso}) are ensemble
averages, which do not necessarily convert to spatial or sample averages.  For a
non-Gaussian field we expect to have to take spatial averages over larger
regions (or bigger samples) than for a Gaussian random field in order to get 
results which approximate equation (\ref{homiso}).

There are various ways we can think about non-Gaussianity. The formal definition
is the departure of the
distribution of the modes from a Normal distribution.  This includes cases
where the phases of the Fourier modes are random and there is no
structure in real space. If we assume the Fourier modes are
independent then homogeneity implies that the modes have random
(uniformly distributed) phases. By the central limit theorem the
distribution rapidly becomes like a Gaussian as we increase the size
of the field and spatial averages rapidly become good estimates of
ensemble averages, i.e. the Fourier modes come close to satisfying
equation (\ref{homiso}). For these cases there will be little or no structure
in real space.

In order
to get structure in the field, we must therefore introduce correlations in the 
modes. This means that spatial averages will take
longer to approach ensemble averages than for a Gaussian random field.  This is
because each realisation of the field has fewer independent modes than
a Gaussian random field does. Therefore, to detect this type of \ng we can test how well
spatial averages obey equation (\ref{homiso}) compared to a Gaussian random
field, rather than looking at the
actual distribution of Fourier modes.
Here we choose to look at $a(\bmath k)$ in a ring of constant $|{\bmath k}|\equiv k$ in
Fourier space, to test the isotropy of spatial averages of the field.


Another reason for looking at the field in Fourier space is to detect
non-Gaussian fluctuations superimposed on a Gaussian background.  
This happens for instance in cosmic string scenarios where the 
effects of strings on photons after last scattering (Kaiser-Stebbins 
effect \cite{KS}) are superimposed on the fluctuations from before last
scattering 
which can be considered to be Gaussian. 
The field may be non-Gaussian on some scales but Gaussian on 
others, making it hard to see the non-Gaussianity in real space.
If there is more power from the non-Gaussian features 
than from the Gaussian on a particular scale this should show in Fourier space.


\section{Another Fourier Transform}

\subsection{Definition}
We can write the Fourier coefficients as
 \begin{equation}
a({\bmath k})=\rho({\bmath k}){\rm e}^{{\rm i}\phi({\bmath k})}
\end{equation}
 
Here we will focus on the amplitudes.  As we are considering constant
$k$ we will write
\begin{equation}
\rho({\bmath k})\equiv\rho_k(\alpha)
\end{equation}
where $\alpha$ is the angle of the direction of $\bmath k$ in polar co-ordinates.

We take another Fourier transform in the angle $\alpha$ around the
ring.
\begin{equation}\label{aft}
F_k(\beta)
=\frac{1}{N}\sum_{n=0}^{N-1} {\rho^2_k(\alpha_n){\rm e}^{-{\rm i}{\alpha_n}{\beta}}}
\end{equation} 

For a completely isotropic field $\langle |a({\bmath k})|^2 \rangle$ in one ring in
Fourier space is 
constant i.e. $\langle\rho^2_k(\alpha)\rangle$ is constant in $\alpha$ so 
$\langle F_k(\beta)\rangle$ is a $\delta$-function in $\beta$.


\subsection{Motivations: Repeated shapes}

This statistic is motivated by the idea of detecting
patterns or shapes that are repeated at different places.  If one
particular feature has a temperature pattern $f(\bmath x)$ with Fourier
transform $\tilde f (\bmath k)$ and the field we are looking at consists of
the feature repeated at $n$ different positions so the field in real
space is
\be
\frac{\Delta T}{T}({\bmath x}) = \sum_{i=1}^n f ({\bmath {x+b}}_i)
\ee
then the Fourier transform
of the whole field is
\be
a({\bmath k}) = \sum_{i=1}^n \tilde f ({\bmath k}) {\rm e}^{{\rm i}{\bmath k} \cdot {\bmath b}_i}
\ee
and the amplitude satisfies
\be\label{norotate}
\rho^2({\bmath k})= |\tilde f ({\bmath k})|^2 \left(n + 2\sum_{i\neq j}\cos( {\bmath k}
\cdot ({\bmath b}_i-{\bmath b}_j))\right)
\ee
The amplitude consists of an envelope defined by the Fourier transform of 
the individual feature together with oscillations  which depend on the
relative positions of the repeated features.
The idea of taking the second Fourier transform is to get a similar 
envelope with oscillations in one dimension around the ring of constant modulus $k$.

As $n$ gets larger $\rho^2({\bmath k}) \rightarrow n|\tilde f ({\bmath
k})|^2$ 
since the cosine averages to zero for many positions, so we 
can extract the Fourier transform of the original feature from 
the Fourier transform of the whole field. We can write this for 
each ring as $\rho_k^2(\alpha) \rightarrow n|\tilde f_k(\alpha)|^2$ 
so we see this applies to each ring also.


But if the shapes are repeated with different orientations, equation 
(\ref{norotate}) becomes
\begin{eqnarray}\label{rotate}
\rho ^2({\bmath k}) &=& \sum_i|\tilde f (R_i{\bmath k})|^2
+ 2\sum_{i\neq j}
\left[\left(\Re e\tilde f(R_i{\bmath k})\Re e\tilde f(R_j{\bmath k}) 
\right. \right. \nonumber \\
& & \left. \mbox{} + 
\Im m\tilde f(R_i{\bmath k})\Im m\tilde f(R_j{\bmath k})\right)
\cos\left( {\bmath k} \cdot ({\bmath b}_i-{\bmath b}_j)\right) 
\nonumber \\
& & \mbox{} +
\left(\Re e\tilde f(R_i{\bmath k})\Im m\tilde f(R_j{\bmath k}) 
\right. \nonumber \\
& & \mbox{} -
\left. \left. \Im m\tilde f(R_i{\bmath k})\Re e\tilde f(R_j{\bmath k})\right)
\sin\left( {\bmath k} \cdot ({\bmath b}_i-{\bmath b}_j)\right) \right]
\end{eqnarray}
where the $\{R_i\}$ are the matrices of the rotations.  
As $n$ gets larger $\rho^2({\bmath k}) \rightarrow \sum_i|\tilde f
(R_i{\bmath k})|^2$ in a similar manner to 
before, which we can write as $\rho_k^2(\alpha) \rightarrow \sum_i
|\tilde f_k (\alpha + \theta_i)|^2$ 
where $\theta_i$ is the angle of the $i$th rotation. It is clear that in each ring this will actually 
tend to a constant (in $\alpha$) as n becomes large if the angles of the rotations are uniformly 
distributed. This is merely a consequence of the increasing isotropy of the field. 

Even with just a few repeated shapes we cannot extract the original
shape as we could in the case 
without rotations without knowing the number and orientations of the
repeated shapes. 
But unless $n$ is very large $\rho_k^2(\alpha)$ should be
distinguishable from a Gaussian.
We know that these features will disappear when averages are taken
over a large enough sample. The
point is to get a statistic which will show the lack of homogeneity
and isotropy on small scales and for small samples.


\subsection{Averaging over small fields}\label{small}
Another way of using this statistic is to look at very small fields
(i.e. those containing
only a few shapes).  To
do statistics on these observations we want to look at many
small fields.  If we say look at several small fields each containing one
shape but at different orientations (as we would expect from isotropy)
then we will get the same Fourier transform rotated each
time. Mathematically it is similar to the calculations leading to
equation (\ref{rotate}) except we don't consider
translations. The real space pattern in the $i$th field is 
 $(\frac{\Delta T}{T})_i({\bmath x}) = \frac{\Delta T}{T}(R_i{\bmath x})$
so the Fourier transform of the field satisfies
 $\rho^2_i({\bmath k}) = \rho^2(R_i{\bmath k})$ which we write in one ring as
\be
\rho^2_{k(i)}(\alpha)=\rho^2_k(\alpha + \theta_i)
\ee
The Fourier transform taken in angle round the ring satisfies
$F_{k(i)}(\beta)= F_k(\beta){\rm e}^{{\rm i}\theta_i\beta}$ so
\be
|F_{k(i)}(\beta)|^2= |F_k(\beta)|^2
\ee
for each field. 

Note that if we averaged over the original
single-Fourier transformed fields $\{\rho^2_{k(i)}(\alpha)\}$ before
taking the second Fourier transform we would get 
\be 
|\sum_{i}F_{k(i)}(\beta)|^2= |F_k(\beta)|^2\left(n + 2\sum_{i\neq j}\cos (\beta
(\theta_i-\theta_j))\right)
\ee
which tends to a constant as we add more fields. We have to take the
second Fourier transform of each field \textit{before} averaging to see
the non-Gaussianity. This is because the features in
$\rho^2_k(\alpha)$ are repeated at different $\alpha$ in each field
and their sum will tend to a constant for many fields.  Taking the
second Fourier transform takes the feature from $\rho^2_k(\alpha)$ and
transforms it to a feature at the same $\beta$ in $F_k(\beta)$ for
each field.

The idea of looking at small fields in this way is an extension of the
idea in \cite{fermag} of analysing several small fields
separately. For the example of a cosmic string on a Gaussian
background used in \cite{fermag} the shape spectrum shows the \ngn
feature well, but the string was always considered to have the same
orientation. If the shape spectrum were averaged over different
directions the result
would disappear. Here we can look at the same fields with different
orientations and the \ng will still be apparent. However, if we
randomly choose small fields from a larger one containing many shapes
we would have to take translations into account as well and the effect
will not be so clear.


\section{Applications}

\subsection{Geometric shapes}
As a simple demonstration of this statistic on an obviously
non-Gaussian field, we will
consider applying it to some regular shapes. These are intended
as mock small scale dust maps. We expect our geometric shapes
 to present
the same type of non-Gaussianity, and obstacles for its
detection, as the actual dust maps. 

First to demonstrate the ideas in Section \ref{small} we will 
look at fields each containing one square at different 
orientations. We first Fourier transform the whole field 
and take the amplitude squared ($\rho^2({\bmath k})$). For 
one square the amplitude is not affected by translation 
of the square. We multiply the field by a Gaussian window to get 
rid of edge effects \cite{hobmag}.

\begin{figure}
 \psfig{file=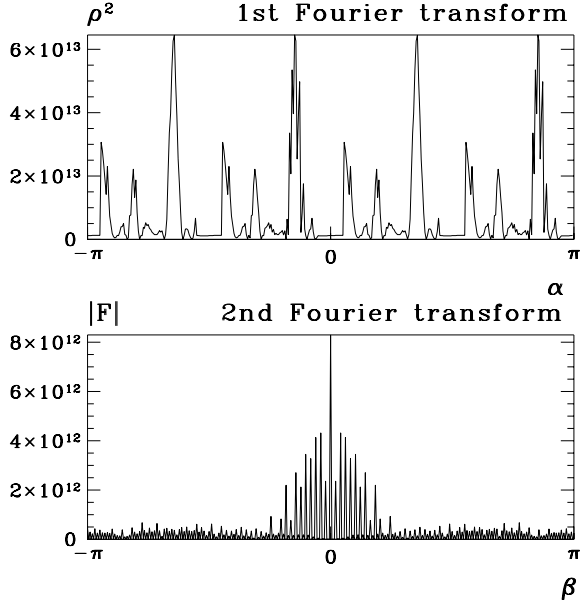,width=8.0cm}
 \caption{First Fourier transform in one ring and modulus of second
 Fourier transform for a field containing 1 square.}
 \label{1square1}
\end{figure}

\begin{figure}
 \psfig{file=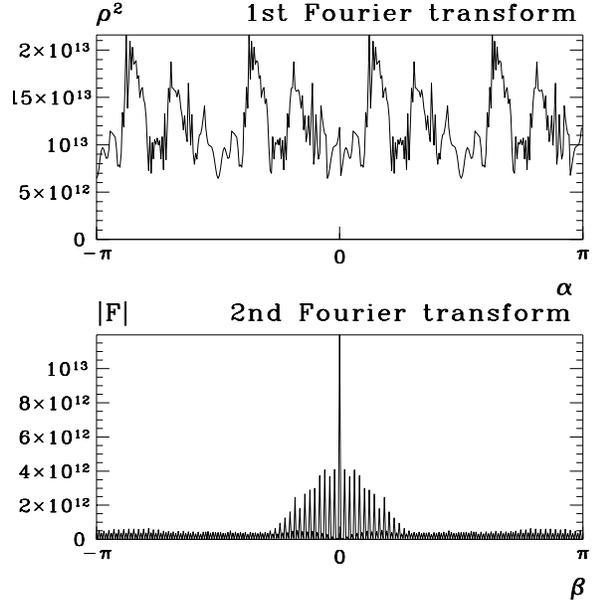,width=8.0cm}
 \caption{Averages of the first and second Fourier transform from 100
realisations of a field containing 1 square.}
\label{1square100}
\end{figure}

In Figure \ref{1square1} we have plotted for one field 
 $\rho^2(\alpha)$ and $|F(\beta)|$ at $k = 60$ from a $200^2$ 
array. This corresponds to a physical scale of $k= 21600/L$ 
where $L$ is the size of the real field in degrees. The y-axis is in
arbitrary units. Figure \ref{1square100} shows
the same functions averaged over 100 fields. The averaged $|F(\beta)|$ is
not in fact identical to $|F(\beta)|$ for one field, because of
numerical effects due to the discrete Fourier transform being on a
lattice. The average settles down quickly, however: $|F(\beta)|$ looks the same
averaged over 10 fields as it does averaged over 100 fields. The first 
Fourier transform, $\rho^2(\alpha)$, 
tends to a constant as we average over more fields as we would expect, but 
is also affected by the lattice.


\begin{figure}
 \psfig{file=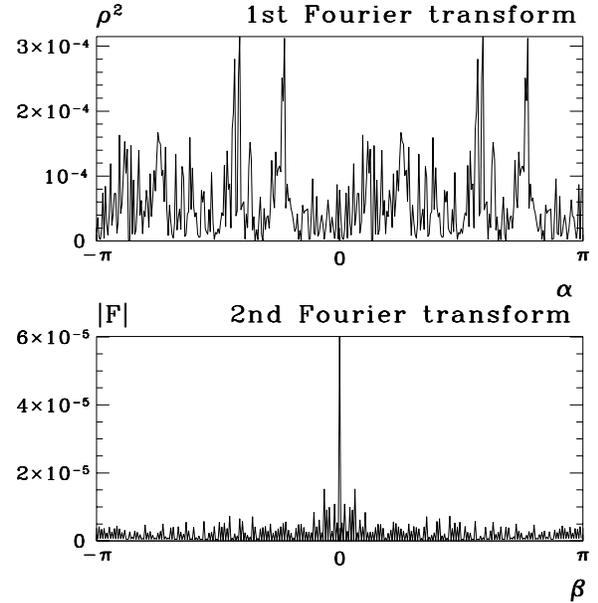,width=8.0cm}
 \caption{First and second Fourier transforms for 1 realisation 
of a Gaussian random field.}
\label{gauss1}
\end{figure}

\begin{figure}
 \psfig{file=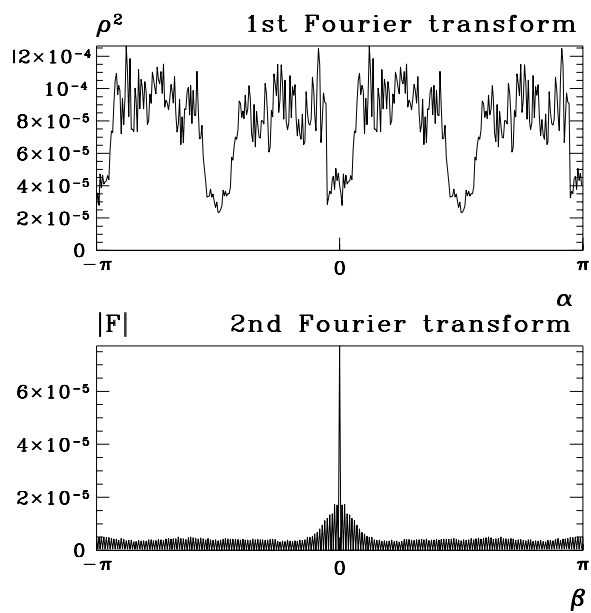,width=8.0cm}
 \caption{Average of first and second Fourier transforms over 100
realisations of a Gaussian random field.}
\label{gauss100}
\end{figure}

Figures \ref{gauss1} and \ref{gauss100} show the equivalent graphs 
for a Gaussian random field. 
Here $|F(\beta)|$ converges rapidly and is distinct from the 
result for squares. The dips in $\rho^2(\alpha)$ are due to the 
window function superimposed in real space to avoid 
edge effects from the first Fourier transform.  The Gaussian is 
treated in this way even though we generate the original map in 
Fourier space so as to model as closely as possible the way in which 
the \ngn map is treated. We would of course expect $|F(\beta)|$ to 
tend to a delta function as the number of fields is increased but 
the window effects spoil this. The window used is in theory circularly 
symmetric but as it is on a lattice it necessarily has 
some squareness in it. The important point is that we get different
results for the Gaussian and \ngn fields.


\begin{figure}
 \psfig{file=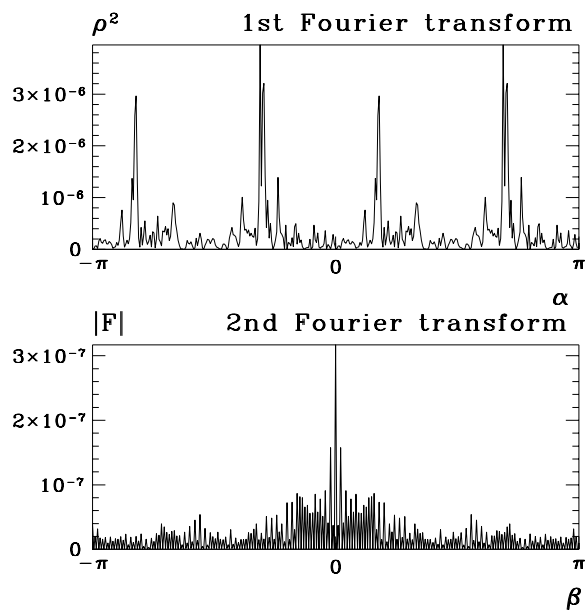,width=8.0cm}
 \caption{First and second Fourier transforms for 10 squares in one field.}
\label{10square1}
\end{figure}

\begin{figure}
\psfig{file=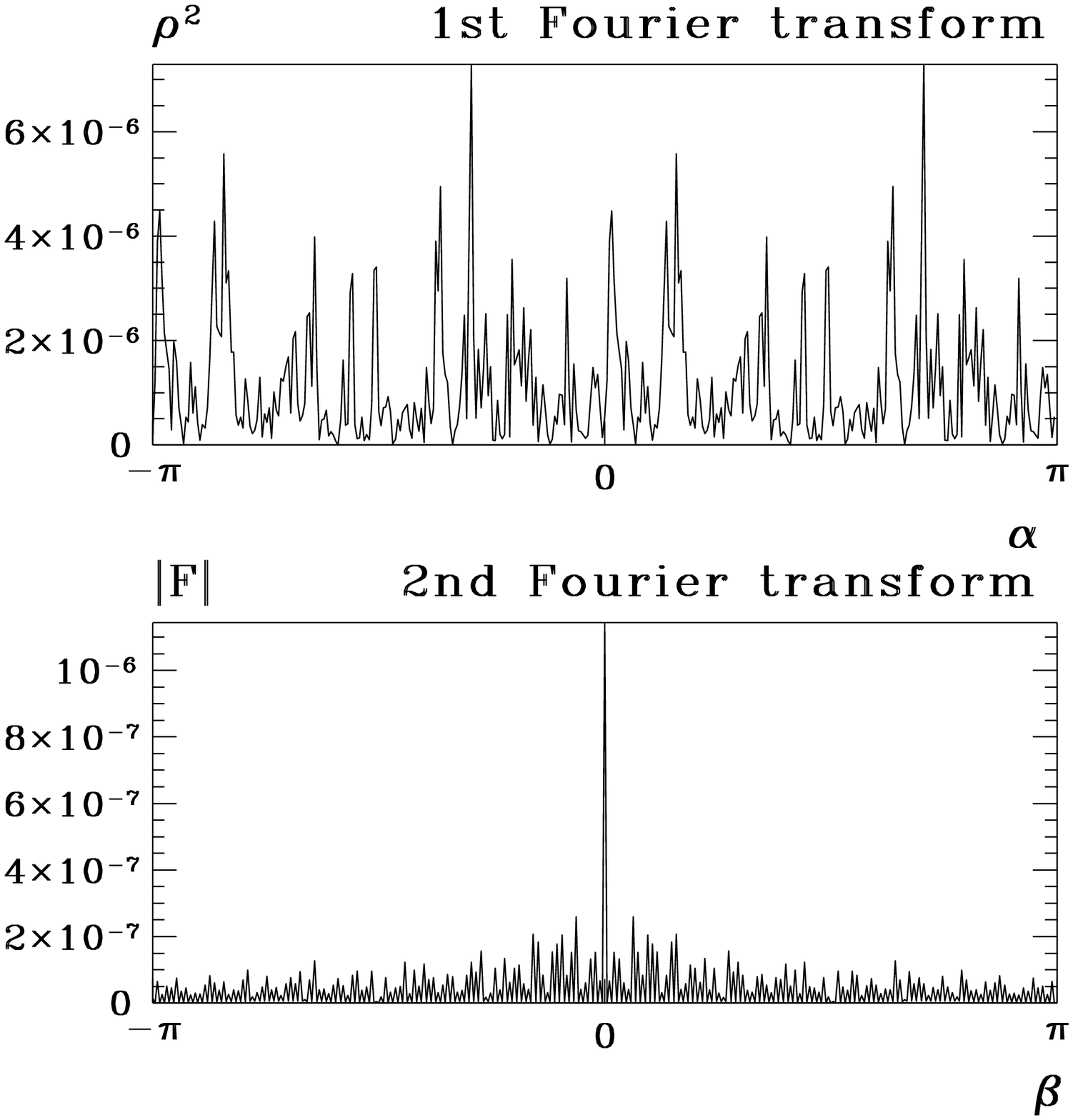,width=8.0cm}
 \caption{First and second Fourier transforms for 100 squares in one field.}
\label{100square1}
\end{figure}

\begin{figure}
 \centerline{\psfig{file=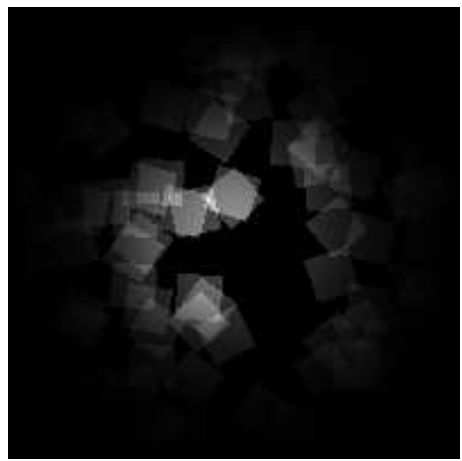,width=6.0cm}\vspace{0.7cm}}
 \caption{100 squares in one field.}
\label{pic100}
\end{figure}

Now we test the statistic on fields with many squares. Figures 
\ref{10square1} and \ref{100square1} show the 1st and 2nd Fourier transforms 
for 10 and 100 squares in one 
field.  We see that as the number of squares increases the result is
not so distinctive. For 100 squares the signal is barely
distinguishable from that of a Gaussian (see Figure \ref{gauss1}). In
fact in real space it is obviously not Gaussian, suggesting this is
not the best way of detecting non-Gaussianity. Figure \ref{pic100} 
shows the field in real space.


\begin{figure}
 \centerline{\psfig{file=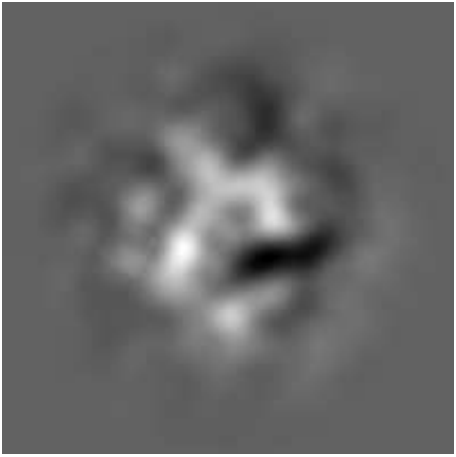,width=6.0cm}\vspace{0.7cm}}
 \caption{1 square superimposed on a field of Gaussian fluctuations.}
\label{pic}
\end{figure}

\begin{figure}
 \psfig{file=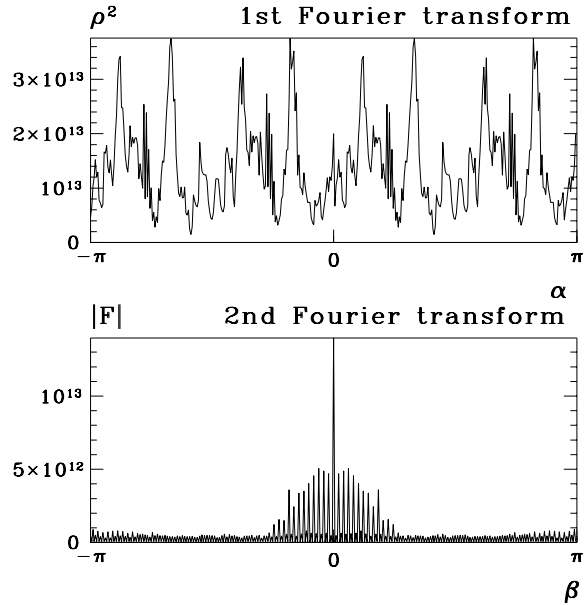,width=8.0cm}
 \caption{Average of first and second Fourier transforms for 10 realisations 
of 1 square superimposed on a
Gaussian background.}
\label{gsq10}
\end{figure}

\begin{figure}
\psfig{file=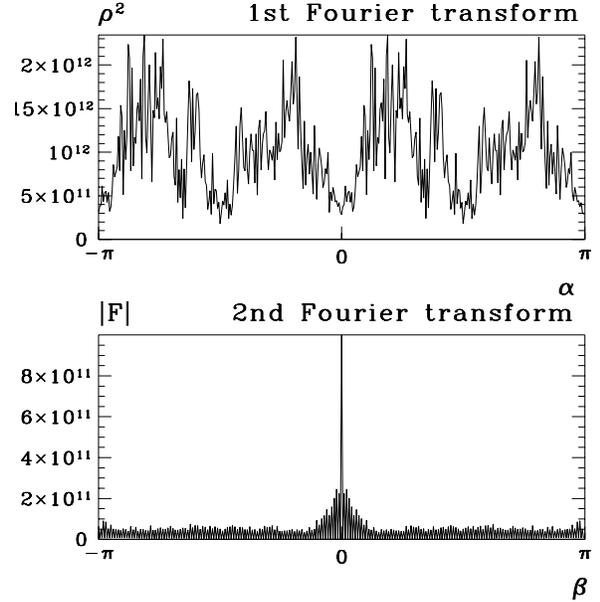,width=8.0cm}
 \caption{Average of first and second Fourier transforms for 10
realisations of a Gaussian.}
\label{gauss10}
\end{figure}

The important test however, is to see if we can detect \ng when it is
superimposed on a Gaussian background so it is not visible in real
space. Figure \ref{pic} shows a real field which is the sum of a
field containing one square and a Gaussian field.  Though in real
space we would not guess there was \ng present, in fact in Fourier
space many rings look different from those produced by Gaussian
fields. An example of one is shown in Figure \ref{gsq10}. This is the
result of averaging over 10 such fields, as in Section \ref{small},
this time with a Gaussian background. Figure \ref{gauss10} shows the
averages for 10 realisations of a Gaussian random field, for comparison.
The characteristic signal of the
square is still present.



\subsection{Cosmic string maps}
Finally we wish to see if this statistic could detect cosmic strings in the
CMB.  The work is motivated by this, since a simple approximation for
a string map is to consider many string segments with random positions
and orientations, and we hope the ideas illustrated above could
apply. We will, however, use the results of string simulations to test
the statistic. 

\begin{figure}
 \psfig{file=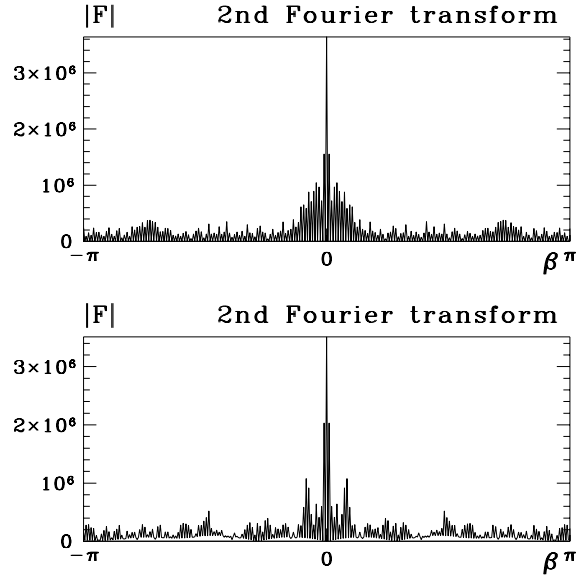,width=8.0cm}
\caption{$|F(\beta)|$ for two different fields containing strings.}
\label{strings}
\end{figure}

The simulation gives us fields which have a physical size of
$2\fdg75$. 
The Fourier transform of the whole
field appears Gaussian, so we 
look at small sections of the simulations to see if the statistic
picks up the non-Gaussianity. We look at fields a quarter of the
size of the simulation. We pick them at random, and analyse them
individually. Figure \ref{strings} shows two graphs of $|F(\beta)|$ obtained
from string maps. Again the y-axis is in arbitrary units.
Many of the graphs obtained look similar to those
obtained from Gaussian fields.

For each string simulation we first obtain $|F^{{\rm NG}}(\beta)|$, then
generate
50 Gaussian fields with the same power spectrum as the string
simulation and treat each one in
exactly the same way as the string maps to obtain $|F^{{\rm G}}_i(\beta)|$,
$i = 1,... ,50$. These
50 values of $|F^{{\rm G}}_i(\beta)|$ are used to obtain the sample 
variance for each $\beta$:
\be
\sigma(\beta) = \sqrt{
\frac{1}{N}\sum_i(|F^{{\rm G}}_i(\beta)|-|\overline{F}^{{\rm G}}(\beta)|)^2}
\ee
where $|\overline{F}^{{\rm G}}(\beta)| = \frac{1}{N}\sum_i|F^{{\rm G}}_i(\beta)|$.

\begin{figure}
 \psfig{file=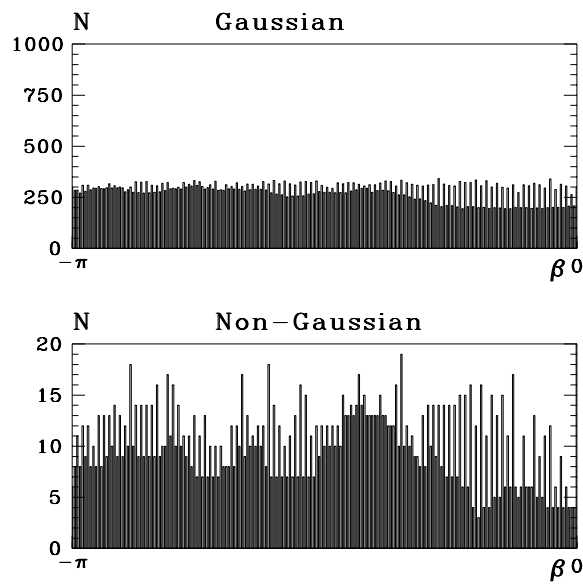,width=8.0cm}
 \caption{Number of simulations giving values of $|F(\beta)|$ outside
$|\overline{F}^{{\rm G}}(\beta)| \pm \sigma(\beta)$ for Gaussian fields 
and fields containing strings.}
\label{hist1}
\end{figure}

Figure \ref{hist1} shows the number of simulations giving a value of 
$|F(\beta)|$ above or below
$|\overline{F}^{{\rm G}}(\beta)| \pm \sigma(\beta)$
 for each $\beta$, for Gaussian
fields and for fields containing strings. The total number
of string simulations is 20 and the total number of Gaussian
simulations is 1000. The axes are scaled so that the proportion of
simulations giving values outside $|\overline{F}^{G}(\beta)| \pm
\sigma(\beta)$ can be
directly compared. Clearly the proportion is higher for the string
simulations, i.e. $|F(\beta)|$ varies more. Large non-zero values of $\beta$ give
more variation in $|F(\beta)|$ than values close to zero.

\begin{figure}
 \psfig{file=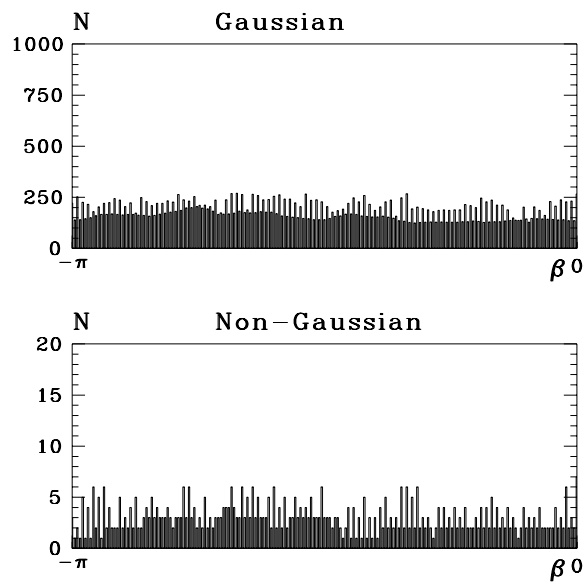,width=8.0cm}
 \caption{Number of simulations giving values of $|F(\beta)|$ outside
$|\overline{F}^{{\rm G}}(\beta)| \pm \sigma(\beta)$ for Gaussian fields 
and fields containing strings
superimposed on Gaussian fields.}
\label{hist2}
\end{figure}

We also want to see if we can detect the strings when they are covered
in real space by Gaussian fluctuations. We have repeated the above
procedure for string simulations superimposed on a Gaussian
field. We have used a power spectrum with an exponential fall off on
small scales as a simple model of the Gaussian fluctuations from
before last scattering: $l(l+1)C_l \sim \exp({-(l/1200)^2})$
with the amplitude of the Gaussian fluctuations being twice that of
the \ngn fluctuations at $l=500$. This power spectrum can be
considered to include noise if it is uncorrelated with the signal.
This time the results are not so clear. Figure \ref{hist2} shows the
same graphs for string simulations superimposed on Gaussian
simulations. Comparing the bottom graph to the top one would not
convince us that it came from a \ngn field.

\begin{figure}
 \psfig{file=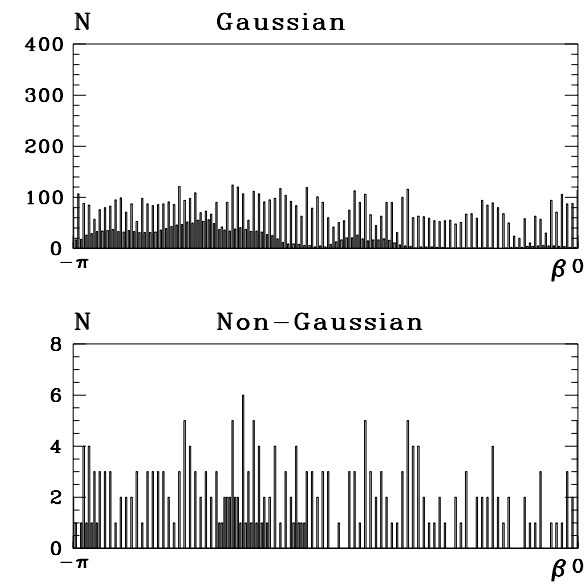,width=8.0cm}
 \caption{Number of simulations giving values of $|F(\beta)|$ below
$|\overline{F}^{{\rm G}}(\beta)| - \sigma(\beta)$ for Gaussian fields 
and fields containing strings
superimposed on Gaussian fields.}
\label{hist3}
\end{figure}

Analysing the data further, however, we can differentiate between the
two cases. Figure \ref{hist3} shows the number of simulations giving
values of $|F(\beta)|$ $\it{below}$ $|\overline{F}^{{\rm G}}(\beta)| - \sigma(\beta)$, for Gaussian
fields, and for
strings and Gaussian superimposed fields. For many values of $\beta$,
the proportion of $|F(\beta)|$ from \ngn fields below
$|\overline{F}^{{\rm G}}(\beta)| - \sigma(\beta)$
is higher than the proportion that we would expect from Gaussian
fields. (This is also true for the strings only case, in fact most of
the time in that case the values of $|F(\beta)|$ which are outside the
`one sigma' range lie below the lower one rather than above the
upper one.)

So to test the sky for \ng we need to look at several fields,
generating Gaussian fields with the same power spectrum for each one, 
and see how many of the fields we are interested in give
values of $|F(\beta)|$ below $|\overline{F}^{{\rm G}}(\beta)| -
\sigma(\beta)$.  Of course we need to look
at a reasonable number of potentially \ngn fields in order to see if
the proportion is more than we would expect from
Gaussian fields.


\section{Discussion}

We have introduced a statistic to look for \ng in the cosmic
microwave background. It is designed to pick out \ngn features which
are superimposed on Gaussian fluctuations and which are therefore not
visible in real space. Since some scales may be more \ngn than others
we choose to separate them out. It is our experience in previous work
\cite{fermag} that in such situations standard statistics, based in
real space, fail to recognize the non-Gaussianity of the signal.

Our statistic is naturally tailored for interferometric experiments, 
which make measurements in Fourier space. Indeed the
algorithm exposed above could easily be turned into a data analysis
package operating over visibilities.

One of the problems with Fourier space statistics is that they recognize
only global shapes. Therefore they become very ineffective when
many individual structures are present. By taking another Fourier transform,
in angle, over modes in a given ring in Fourier space, we can factor
out orientations, and be only sensitive to an average shape of
individual structures. Their random positions, on the other hand,
will eventually make our statistic very ineffective as the number of
objects becomes large. We found
that although we have improved on previous work, still there is a
limit on how many individual structures there may be in the field
before the statistic fails to pick out their non-Gaussianity.

Finally, some comments are in order regarding the applications we have
considered. The Gaussian component we have considered may include
noise, if it has the right power spectrum. Hence we have already
considered the effects of a simplified form of noise: it will merely
reduce the band of scales where the non-Gaussian signal dominates, 
typically providing it with an upper boundary. An exception is the case
of non uniform noise in the u-v plane, present in most interferometer
experiments. Non uniform noise in the Fourier domain will look 
non-Gaussian, for it is a Gaussian process which is not isotropic
or translationally invariant. Therefore an extra element of confusion,
not dealt with in this paper, will appear.

Regarding the non-Gaussian component we have considered two types
of signal. 
We have looked at some of the properties of the statistic when it
is applied to a field containing one or more distinct shapes.
The geometrical constructions presented are somewhat reminiscent
of the jagged structures present in small scale dust maps.
Experimenting on fields containing many structures we saw our statistic
rapidly become blind to their non-Gaussianity. However interferometer
fields are often small. They would therefore contain only a few, but
more than one, of these structures. This is something our statistic
can cope with, unlike previous Fourier space based statistics.

We have also studied our statistic when applied to cosmic string maps.
We found that some individual small fields showed \ng in the signal. 
If the results were averaged over many fields the effect of \ng 
on $|F(\beta)|$ would not show. Once again this strategy is ideal
for interferometers, for which combining small fields into a large
field is in fact a rather awkward operation.

\section*{Acknowledgements}
We would like to thank Tom Kibble for useful discussion and Pedro
Ferreira for the string simulation code. A.A. and J.M. would like to 
thank the CfPA at Berkeley, where we were
visiting when this project was started.
A.L. is supported by PPARC and J.M. by a Royal Society University Research
Fellowship.



\begin{thebibliography}{99}

\bibitem[Albrecht 1996]{coher}Albrecht A. 1996, Coherence and Sakharov Oscillations in the
Microwave Sky, in Proceedings of the XXXIst Recontre de
Moriond, 'Microwave Anisotropies'.
\bibitem[Albrecht, Battye \& Robinson 1997a]{abr}Albrecht A., Battye R. A. and
Robinson J. 1997a, \prl, 79, 4736
\bibitem[Albrecht, Battye \& Robinson 1997b]{abr2} Albrecht A., 
Battye R. A. and Robinson J. 1997b, \prd, submitted
\bibitem[Albrecht et al. 1998]{abrw}Albrecht A., Battye R. A.,
Robinson J. and Weller J. 1998, in preparation
\bibitem[Avelino et al. 1997]{shellard}Avelino P. P., Shellard E. P. S., Wu J. H. P. and
Allen B. 1997, preprint
\bibitem[Bardeen 1980]{bardeen}Bardeen J. M. 1980, \prd, 22, 1882
\bibitem[Bardeen et al. 1986]{bardstat}Bardeen J. M., Bond J. R., Kaiser N. and
Szalay A. S. 1986, \apj, 304, 15
\bibitem[Battye 1997]{richard}Battye R. A. 1997, \prd, 55, 7361
\bibitem[Battye, Robinson \& Albrecht 1997]{abr3}Battye R. A., Robinson
J. and Albrecht A. 1997, preprint
\bibitem[Bond \& Efstathiou 1987]{bondefst}Bond J. R. 
and Efstathiou G. 1987, \mnras, 226, 655
\bibitem[Bouchet, Bennett \& Stebbins 1988]{phases}Bouchet F. R., Bennett D. P. and Stebbins A. 1988,
\nat, 335, 410
\bibitem[Durrer \& Sakellariadou 1997]{durrer}Durrer R. and 
Sakellariadou M. 1997, \prd, 56, 4480
\bibitem[Ferreira \& Magueijo 1997]{fermag}Ferreira P. G. and Magueijo J. 1997, \prd, 55, 3358
\bibitem[Ferreira, Magueijo \& Silk 1997]{cumul}Ferreira P. G., 
Magueijo J. and Silk J. 1997, \prd, 56, 4592
\bibitem[Hobson \& Magueijo 1996]{hobmag}Hobson M. P. and 
Magueijo J. 1996, \mnras, 283, 1133
\bibitem[Hu \& Sugiyama 1995a]{HS1}Hu W. and Sugiyama N. 1995a, \apj, 444, 489
\bibitem[Hu \& Sugiyama 1995b]{HS2}Hu W. and Sugiyama N. 1995b, \prd, 51, 2599
\bibitem[Hu, Sugiyama \& Silk 1997]{nature}Hu W., Sugiyama N. and
Silk J. 1997, \nat, 386, 37
\bibitem[Kaiser \& Stebbins 1984]{KS}Kaiser N. and Stebbins A. 
1984, \nat, 310, 391
\bibitem[Kibble 1980]{Kib}Kibble T. W. 1980, \physrep, 67, 183
\bibitem[Kogut et al. 1996]{Kogut}Kogut A., Banday A. J., Bennett C. L., Gorski K.,
Hinshaw G., Smoot G. F. and Wright E. L. 1996, \apj, 464, L29
\bibitem[Liddle \& Lyth 1993]{lidlyth}Liddle A. R. and
Lyth D. H. 1993, \physrep, 231, 1
\bibitem[Luo 1994]{XLuo}Luo X. 1994, \prd, 49, 3810
\bibitem[Magueijo et al. 1996]{mafc}Magueijo J., Albrecht A., 
Ferreira P. G. and Coulson D. 1996, \prd, 54, 3727
\bibitem[Pando \& Fang 1996]{pando}Pando J. and Fang L. Z. 1996,
preprint
\bibitem[Peebles 1970]{pyu}Peebles P. J. E. and Yu J. T. 1970, \apj,
162, 815
\bibitem[Pen, Seljak \& Turok 1997]{neil}Pen U. L., Seljak U. and
Turok N. 1997, \prl, 79, 1615
\bibitem[Perlmutter et al. 1998]{sn1}Perlmutter S. et al. 1998, \nat,
391, 51 
\bibitem[Schmalzing \& Gorski 1998]{gorski}Schmalzing J. and Gorski K. M. 1998, \mnras, in press
\bibitem[Smith \& Vilenkin 1987]{av}Smith A. G. and Vilenkin A. 1987, \prd, 36, 990
\bibitem[Steinhardt 1995]{cosmcross}Steinhardt P. J. 1995, in Particle and
Nuclear Astrophysics and Cosmology in the Next Millenium, 
ed. E. W. Kolb and R. Peccei (Singapore: World Scientific) 51 
\bibitem[Turok 1996]{mimic}Turok N. 1996, \prl, 77, 4138
\bibitem[Vachaspati \& Vilenkin 1984]{vv}Vachaspati T. 
and Vilenkin A. 1984, \prd, 30, 2036
\bibitem[White, Scott \& Silk 1994]{1.3}White M., Scott D. and 
Silk J. 1994, \araa, 32, 319
\bibitem[Winitzki \& Kosowsky 1997]{sergei}Winitzki S. and Kosowsky A. 1997, New Astronomy,
Vol. 3, No. 2, 75



\end{thebibliography}
\end{document}